# DISCRETE EVENT SIMULATION FOR PORT BERTH MAINTENANCE PLANNING

Ruqayah Alsayed Ebrahim
Shivanan Singh
Yitong Li
Wenying Ji

Department of Civil, Environmental, and Infrastructure Engineering
George Mason University
4400 University Drive
Fairfax, VA 22030, USA

## ABSTRACT

Industrial and commercial ports, which are one of the three main hubs to the country, require 24/7 operations to maintain the goods export and import flow. Due to the aging and weather factors, berths require regular maintenance, such as replacing old piles, timber finders, marine ladders, rubber fenders, and deck slabs. For efficient berth maintenance, strategies are highly desired to minimize or eliminate any delays in operations during the maintenance. This paper develops a discrete event simulation model using *Simphony*.NET for berth maintenance processes in Doha Port, Kuwait. The model derives minimum maintenance duration under limited resources and associated uncertainties. The model can be used as a decision support tool to minimize interruption or delays in the port maintenance operations.

## 1 INTRODUCTION

The construction industry is complex and requires detailed analysis for enhanced overall performance. Berth maintenance/rehabilitation plays an important role in port or harbor projects, and it is deemed as an infrastructure project involving numerous risks compared to other construction projects. Examples of these risks include shortage of materials, extreme weather (e.g., dust storms), and equipment breakdowns. To ensure the efficiency of berth maintenance projects, strategies that consider resource limitations and risks need to be developed. This research focuses on the berth maintenance project in Doha Port, Kuwait. The small, shallow, and merchant port is located in the 'S' part of Kuwait servicing dhows, barges, and other coastal vessels operating between ports of the Persian Gulf. The semi-closed basin design, as shown in Figure 1, handles 7,000 small vessels and approximately 200,000 tons of cargo annually (Shipnext 2022). Recently, Kuwait Ports Authority has pursued its goal of modernizing and expanding its ports' facilities to accommodate larger vessels and reduce discharge times of cargo. Therefore, site investigations and tests were conducted on the port docking facilities (Figure 1) to investigate the durability of the old concrete deck slabs, and their capacity to withstand large and heavy loads from present-day cargo.

The tests identified a major problem with the berth—the piles and slab decks were completely damaged. Due to the damage, the berth was insufficient to receive ships, which have caused delays in the operation of exporting and importing goods. As such, it was required that significant rehabilitation activities should be conducted. Precast construction was employed (specifically precast concrete slabs for the slab decks) in an attempt to accelerate the project duration and minimize further delays.

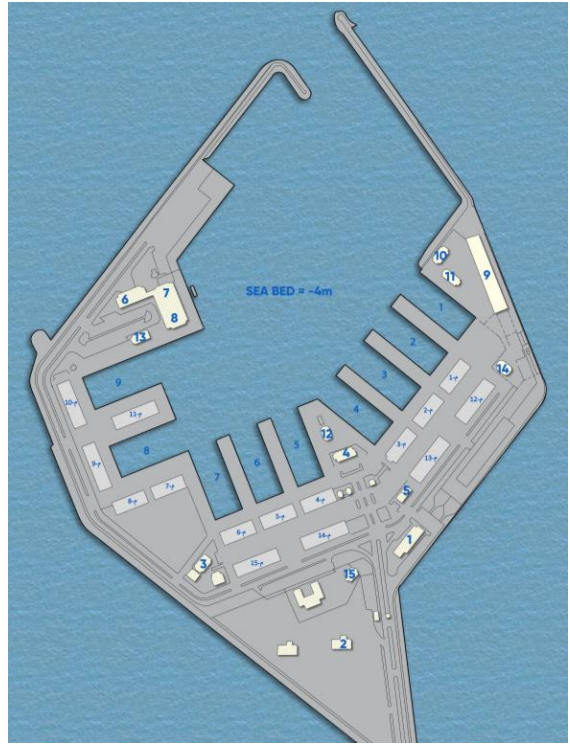

Figure 1: Kuwait Ports Authority, Doha Port Layout, Kuwait (Scale 1:5,000).

The construction process involves 12 phases (shown in Figure 2): removal works, reinforcement, pouring concrete, structure pile installation, fender pile installation, reinforcement of precast concrete slabs, precast concrete slab installation, front deep beam reinforcement, front deep beam concrete pouring, timber fender installation, and rubber fender installation and painting.

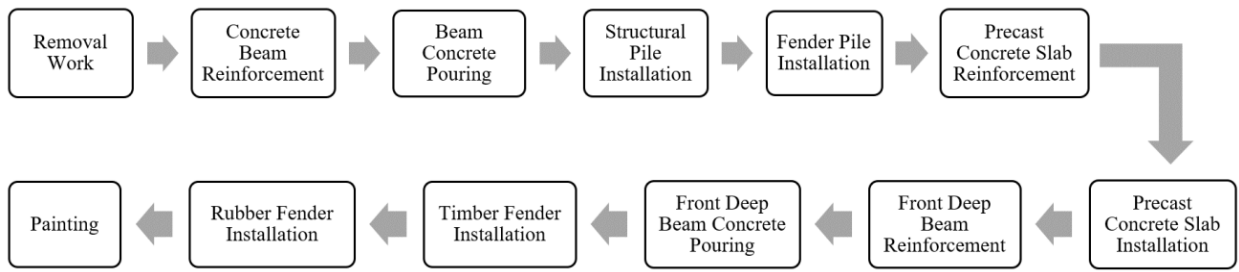

Figure 2: Phases of the berth maintenance project.

In detail, "*removal work*" starts with the removal of the existing 23,000 $m^2$ concrete deck slabs and the removal of the outer concrete layer of the concrete beam until they reach the reinforcement layer. If the reinforcement is not completely damaged, then it will be treated with Epoxy, otherwise, it will be replaced by new reinforcement. Then, the treatment of pile caps starts followed by "*concrete beam reinforcement*" and "*beam concrete pouring*." Prestressed "*structural pile installation*" starts where they dug around 60 structural piles (4 piles/day) using a 3 ton hammer and a crane. The structural pile is 36 cm (about 1.18 ft) × 36 cm (about 1.18 ft) and 16 meters long (size $2.07 m^3$). Then, prestressed "*fender pile installation*" starts where they dug around 108 fender piles (5 piles/day). The fender pile is 36 cm × 36 cm and 17.5 meters long. At an offsite facility, *"precast concrete slab reinforcement"* is installed in a suitable steel mold, and concrete is poured. Once the concrete has acquired an adequate compressive strength, slabs are then transported to site where 23,500 $m^2$ of *"precast concrete slab installation"* is performed with a crane followed by *"front deep beam reinforcement"*, *"front deep beam concrete pouring"*, *"timber fender installation"*, *"rubber fender installation"* and *"painting"*, respectively. Table 1 shows the relationship between resources (materials, labor, and equipment) and project activities.

The objective of this research is to develop a simulation model to mimic the berth maintenance process in Doha Port, Kuwait, using the construction simulation platform *Simphony*.NET (AbouRizk et. al. 2016). This platform was chosen because it was specifically developed for the construction industry and the variety of modeling elements can realistically mimic what occurs on construction sites. In detail, the proposed discrete event simulation model incorporates the process of the berth maintenance project as well as resource limitations and uncertainties associated with the process. The model can be used to inform owners, project managers, and contractors with adequate resource allocation and optimum scheduling performance. Furthermore, the client can use this model to realistically criticize construction proposals submitted by contractors during the bidding process.

Table 1: Project resources associated with their tasks.

| **Resources** | **Tasks** |
|---|---|
| Crane | Loading and unloading the following materials: concrete formwork, piles, precast concrete slabs, timber fenders and rubber fenders |
| Excavator | Excavation |
| Jackhammer | Old concrete deck slabs removal |
| Pile hammer | Installing piles |
| Dump Trucks | Hauling excavated material |
| Concrete Pump | Pouring concrete |
| Concrete Trucks | Loading concrete to the site |
| Forklift | Loading and unloading timber and rubber fenders |
| Piling Crew | Pile head cap treatment, inner piling and outer piling |
| Concrete Crew | Reinforcement installation, concrete pouring and concrete curing |
| Laborers | Old concrete deck slab removal, epoxy treatment, timber and rubber fender installation and painting |

## 2 LITERATURE REVIEW

Berth maintenance/rehabilitation process depends on the tests and site investigations results, and the size of the damage (e.g., minor or major damage). In cases of minor damage, the maintenance/rehabilitation process begins with the repair of areas with spalling and exposed reinforcement. Then, concrete pavement repair work begins by sealing cracks followed by installing new bollards, fenders, life ladders, and manholes. In cases of major damage, the process includes deck slabs repair, pile cap repairs, demolition and removal of existing infrastructure, dredging and dumping of material, reclamation of land, construction of embankments, construction of reverted slopes, provision of infrastructure utilities (water, electricity, etc.), erection of reefer stacks, and construction of drainage systems (KPA 2022).

Seaborne trade development is affected by demand trade patterns and competition. The difference in demand could create an uncertain environment to develop a strategic decision-making process. "Strategic and tactical decisions tend to be capital intensive and must be flexible to adapt and expand in terms of infrastructure and technological changes in the long run" (Stopord 2008). Tools that have been used to identify and visualize the process of berthing and port operations are Business Process Modeling Notation (BPMN) (BPMI 2004) and Discrete Event Simulation Modeling (DES) (Caceres et. al. 2015). Another tool called STAADPRO (Research Engineers international 1997) has been used to model the design of a proposed marine berthing structure using induced load distribution (Vivek and Prasad 2016). No research has been found about simulation models of berth maintenance/rehabilitation processes involving risks/uncertainties that answers the question "what if?"

*Simphony.NET* construction simulation appeared to be the most feasible as it can model complex processes involving uncertainties. As such, understanding of such processes can be communicated and validated by experts and then can be analyzed by us to look for alternatives that can be easily translated into execution language by non-technical users (Hajjar and AbouRizk 1999). Simulation modeling of berth maintenance/rehabilitation process is a digital prototype of a real-world problem to predict the performance by presenting materials, equipment, and labor as resources, operations as tasks, and equipment breakdowns or severe weather conditions as sub models. The simulation model presented in this paper analyzes the construction process of rehabilitating the berth taking into consideration of uncertain events that can affect the project schedule and its outcome. The results of such a model can help the owner/contractor/project manager control and monitor the project progress, schedule, and cost.

Time-cost tradeoff projects or problems in project management can involve some crashing. Crashing in construction projects include activity crashing due to materials, labor, and equipment crashing, which means adding costs to meet the project schedule (Pena 2009). To minimize the total project cost and time, the crashing of an activity answers the question "what is the maximum number of time units that an activity can have during crashing and use it to reduce the activity time?" (Pena 2009). By identifying and providing $z$ input parameters such as the number of activities, the probability distribution of completion of each task or activity, the maximum completion time, and the maximum crashing time, a simulation model can be constructed to determine the average project cost/time resulting from crashing.

## 3 METHODOLOGY

The presented methodology consists of one main component of a simulation model using the construction simulation platform, *Simphony*.NET. The rehabilitation of a berth follows a particular sequence of steps determined by engineers to produce an effective, structurally sound final product. In this scenario, the old berth's concrete components needed to be partially demolished and then repaired, followed by new construction. The model developed by *Simphony*.NET follows the exact sequencing of events in a controlled environment. The entire model can be simplified to smaller activities where various resources were determined and configured to be captured or released as required. These resources move from one activity to another, however, are not shared between them simultaneously. Due to site space constraints (being located next to the ocean), access is restricted, thereby discouraging concurrent construction activities.

Understanding previous berth construction provided a basis in which we can develop the model, and further change resource elements to improve production. As such, the model was developed in a way to closely mimic what occurs on site. Entities are not flowing freely from the beginning to the end of the model, but rather grouping intermittently to finish a task, and then moving to other tasks at similar times. This modelling behavior was purposely adopted since on-site activities occur the same way. For example, all exposed steel must be completely treated (grouped) with epoxy before new reinforcement installation can occur. Similarly, all concrete for the deck needs to be poured together (completely grouped) before moving onto the curing phase. With this modelling configuration, the client or engineer can more accurately simulate construction events, thereby allowing them to choose the most suitable resource allocation strategy.

## 4 CASE STUDY AND RESULTS

### 4.1 Data Preparation and Model Inputs

On November 10, 2019, rehabilitation of the first berth commenced and lasted 189 days. Compared to the contractor's bid schedule of 156 days, there was a delay of 33 days which ultimately increased project costs.

This also delayed docking of container ships which resulted in a penalty fee that the port authority ultimately had to pay. The actual completion times of each activity for one berth (100m in length) are given in Table 2.

Table 2: Duration of rehabilitation project tasks.

| Project Task No. | Project Task | Time (days) | Project Task No. | Project Task | Time (days) |
|---|---|---|---|---|---|
| 1 | Jackhammering Concrete | 30 | 11 | Piling (Inside Berth) | 15 |
| 2 | Excavation | 10 | 12 | Piling (Outside Berth) | 22 |
| 3 | Hauling Material Offsite | 4 | 13 | Install Precast Slabs | 2 |
| 4 | Applying Epoxy to Steel | 3.25 | 14 | Install Deck Reinforcement | 8 |
| 5 | Treating Pile Heads | 30 | 15 | Pour Deck Concrete | 2 |
| 6 | Install New Reinforcement | 24 | 16 | Cure Deck Concrete | 3 |
| 7 | Install Formwork | 15 | 17 | Install Timber Fenders | 8 |
| 8 | Pour Beam Concrete | 3 | 18 | Install Rubber Fenders | 10 |
| 9 | Cure Beam Concrete | 3 | 19 | Painting | 2 |
| 10 | Remove Formwork | 10 | | | |

## 4.2 Model and Assumptions

The model was built to closely mimic the activities of construction of a previous berth. An event must be started and finished in order to move on to the other event, and so on. The end of the model would signify the completed construction of one 100m long berth. The resources linked to the main model are shown in Figure 3. The main model schematic is illustrated in Figure 4. Explanations of the model elements are given in Table 3.

Table 3: Model elements' legend.

| Element | Description | Element | Description |
|---|---|---|---|
| 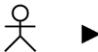 Resource | Represents a resource with a specified number of servers | 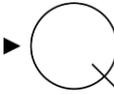 File | A file in which entities wait for resources |
| 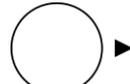 Create Entity | Creates entities | 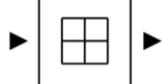 Batch Entities | Batches a group of entities |
| 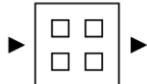 Unbatch Entity | Allows an entity to take one of two paths depending on a specific condition | 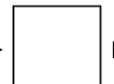 Task | Delays an entity for a specified amount |
| 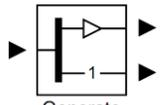 Generate | A modeling element that generates clones of an entity and sends them out at a separate output point | 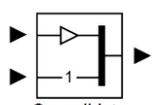 Consolidate | A modeling element that consolidates a specified number of entities before releasing one |
| 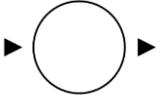 Function | Generates or consolidates entities | 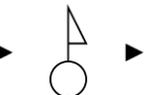 Counter | Counts the number of entities passing through the element |
| 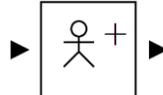 Capture Resource | Allows an entity to request servers of one or more resources | 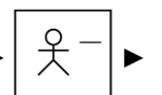 Release Resource | Allows an entity to release servers of one or more resources |
| 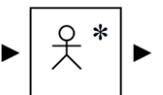 Preempt Resource | Allows an entity to preempt a single server of a resource | 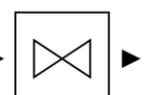 Valve | Allows entities to pass or block them depending on a state variable |
| 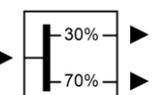 Probabilistic Branch | Allows an entity to take one of multiple paths depending on specified probabilities | 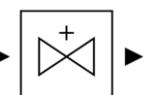 Activator | Allows an entity to change the state of a valve |
| 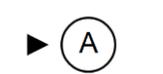 | Receive entities | 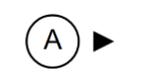 | Send entities |
| 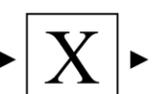 Execute | Executes a formula when an entity passes through | 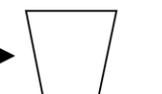 Destroy | Destroys entities |

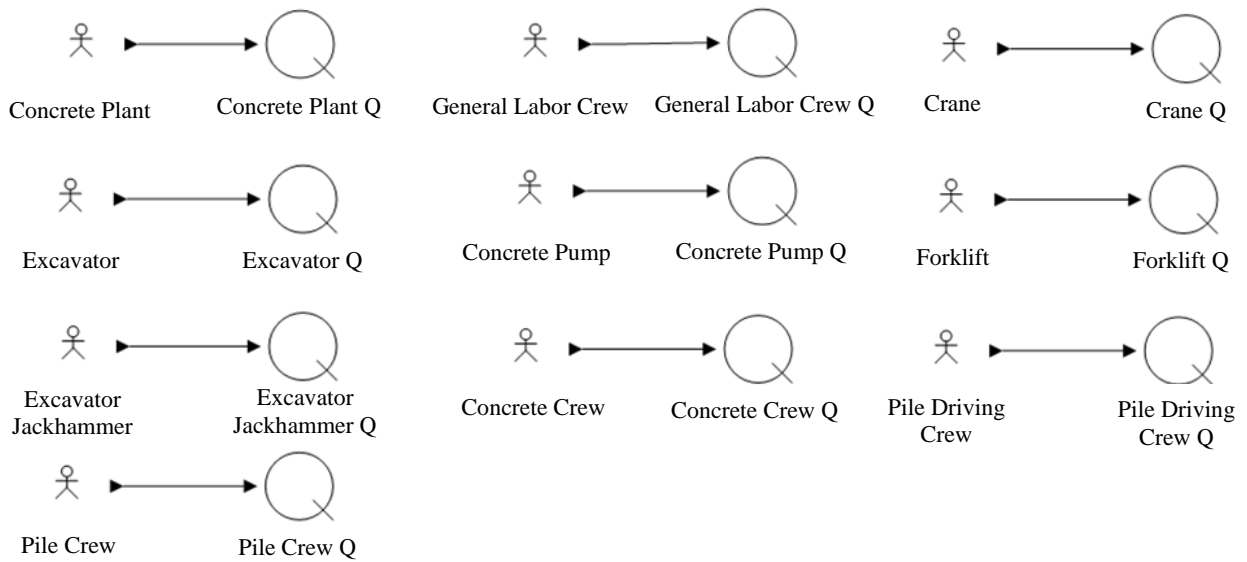

Figure 3: Resources used for modeling the project using *Simphony*.NET.

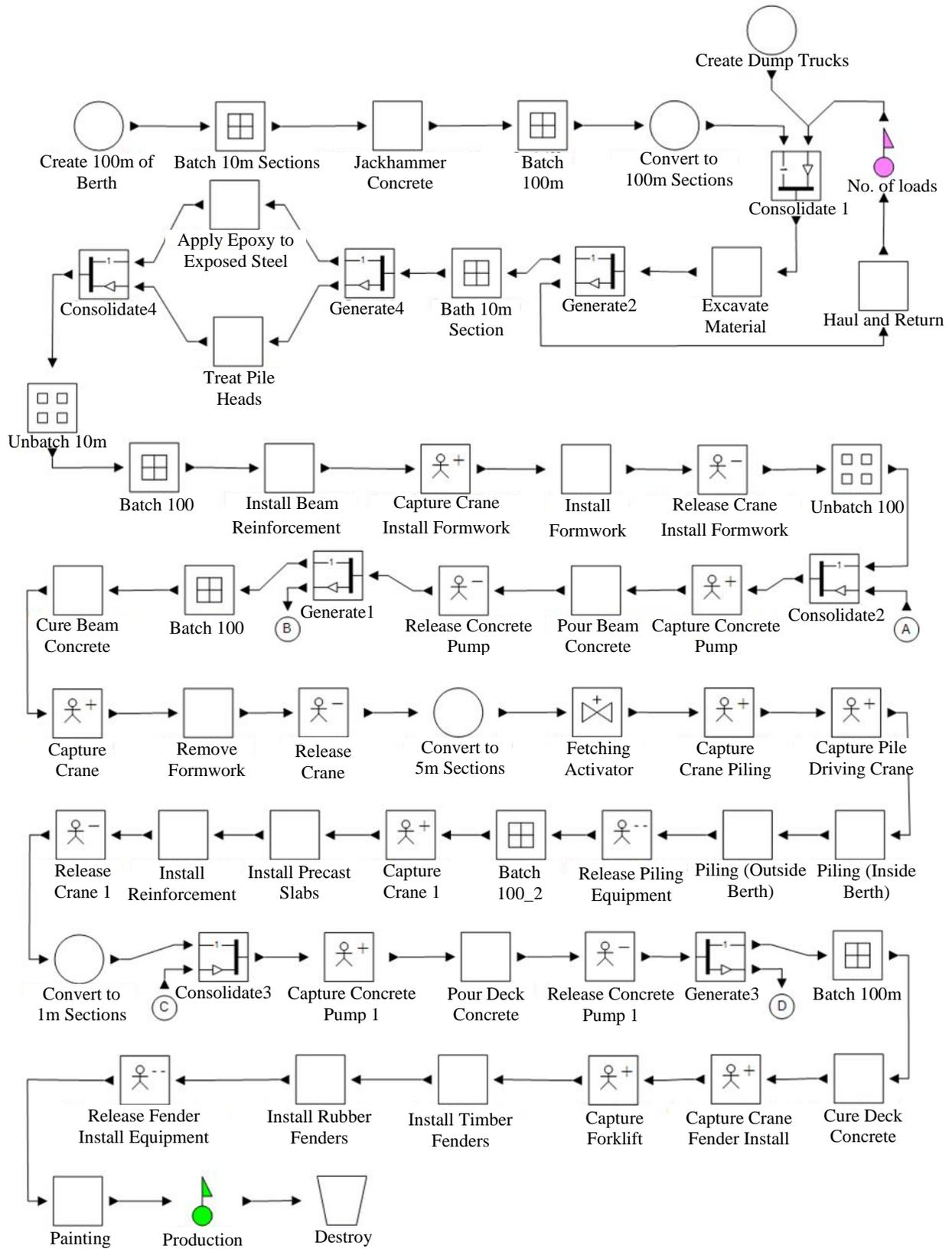

Figure 4: Simulation base model of the project using *Simphony*.NET.

It was assumed that the *General Labor Crew* resource can be utilized for several activities throughout the project. However, specialized activities like concreting and pile driving required a separate specialized crew. Also, due to site space restrictions and safety, numerous types of work cannot occur simultaneously, as such resources must remain idle until their respective activity is ready to be conducted.

Three sub-models, shown in Figure 5, were created to complement the functions of the main model, but at offsite locations. They include the following: (1) concrete for beam; (2) concrete for deck; (3) fetching piles.

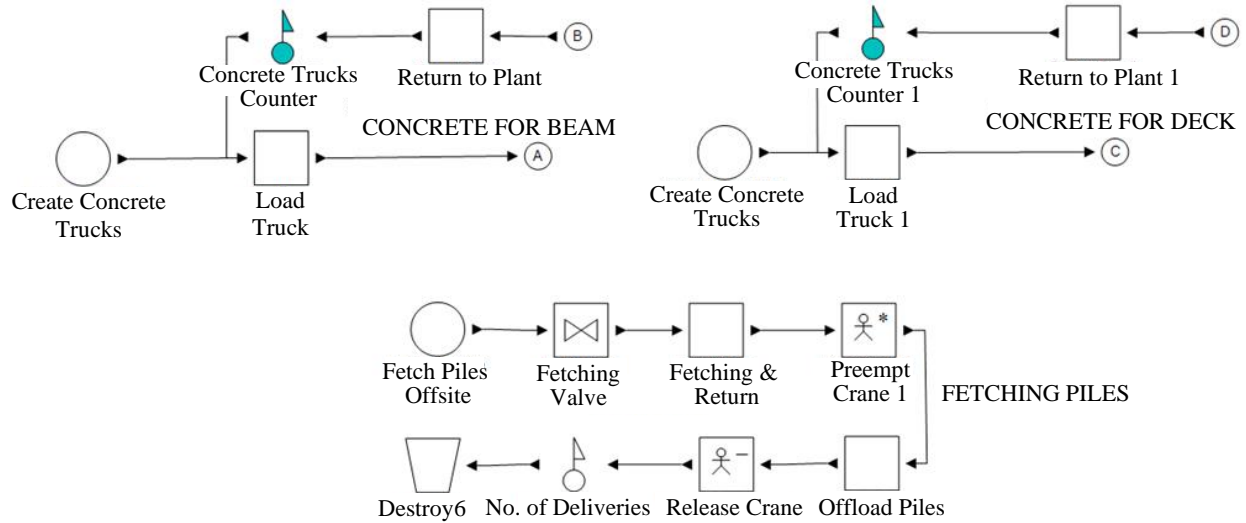

Figure 5: Simulation sub-models.

Furthermore, three other sub-models shown in Figure 6 were created to mimic weather complications and equipment breakdown. They include the following: (1) bad weather; (2) crane breakdown; (3) jackhammer breakdown.

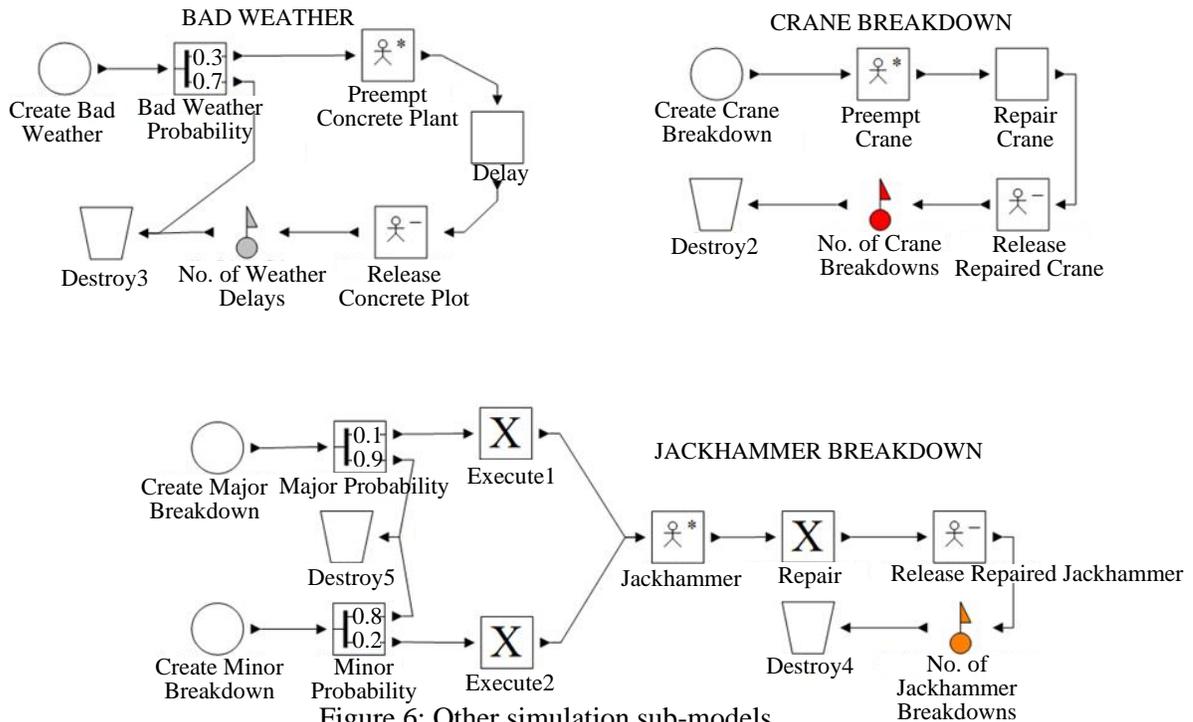

Figure 6: Other simulation sub-models.

### 4.3 Results

The *Simphony*.NET model was run following the sequencing of actual events as recorded by project management. The equipment and labor resources were allocated to each activity based on the contractor's proposal. The activity time was specified in the model with reference to actual performance, and not the proposed schedule. Furthermore, different breakdowns and uncertainties were introduced to determine how the project cycle was affected. The results acquired are summarized in Table 4.

Table 4: *Simphony*.NET simulation results with uncertainties.

| Scenario (Cumulative) | Production Rate | Total Production Time | Comments |
|---|---|---|---|
| Ideal Conditions | 0.52 m/day | 193.38 days | All tasks can be conducted without interruption |
| Bad Weather | 0.51 m/day | 195.38 days | 30% chance that there is bad weather every 10 days |
| Crane Breakdown | 0.47 m/day | 215.00 days | Every 5 days the crane breaks down and needs repair |
| Jackhammer Breakdown | 0.43 m/day | 233.46 days | Jackhammer experiences minor and major breakdowns |

The simulated result for the 'Ideal Conditions' scenario was similar to actual construction performance in Kuwait. As shown in Table 4, whenever uncertainty regarding weather conditions or equipment breakdown was introduced in the model, the subsequent delays experienced caused decreases in production rate from 0.52 m/day to 0.43 m/day and increases in total production time from 193.38 days to 233.48 days. These uncertain events are commonly seen in construction projects. However, they may not be easily incorporated into a bidding schedule even though they are certainly realistic. To mitigate these uncertainties, modifications of available resources can be conducted to offset the effects of delays. As such, resources in the model were tweaked to offer viable solutions whilst being kept within a reasonable project time frame. The results acquired are summarized in Table 5.

Table 5: *Simphony*.NET simulation results with all uncertainties and resource modification.

| Scenario (Cumulative) | Production Rate | Total Production Time | Comments |
|---|---|---|---|
| With All Uncertainties | 0.43 m/day | 233.46 days | All breakdowns and bad weather |
| Additional Concrete Trucks | 0.47 m/day | 211.21 days | Trucks increases from 2 to 8 trucks |
| Additional Jack Hammers | 0.52 m/day | 191.46 days | Jackhammers increased from 1 to 3 |
| Additional General Labour | 0.59 m/day | 170.00 days | General labor crew increased from 1 to 2 |
| Additional Concrete Crew | 0.60 m/day | 166.71 days | Concrete crew increased from 1 to 2 |
| Additional Concrete Pump | 0.62 m/day | 161.71 days | Concrete pump increased from 1 to 2 |

Table 5 shows enhanced production due to improved resources. With additional concrete trucks, jack hammers, general labor and concrete crews, and an additional concrete pump, completion of the berth can be reduced from 233.46 days to 161.71 days, which is similar to the project schedule bid by the contractor. Consequently, the production rate increased from 0.43m/day to 0.62 m/day.

The total project cost was $773,422 for the maintenance of 1 berth ($17,633,191.329 for 7 berths) including the delays with no additional equipment or labor. Additional equipment and labor (including crew and engineers) will increase the cost to $971,872 - $1,368,772. An example of control can be observed when the jackhammer experiences a major breakdown, thereby requiring 7 days for repair/replacement. In such an event, project managers can determine from the simulation model that having other jackhammers working simultaneously can reduce the length of delay caused by a 7-day absence of the broken jackhammer. Therefore, the downtime and resources are being controlled to mitigate against project schedule growth. Another example of control relates to increased placement rate of concrete with additional pumps and trucks, thereby saving on the entire construction schedule. In the end, project managers would have to make the final decision to allow the operation of additional resources. Depending on the sensitivity of port operations during construction, it may be more feasible to bolster project finances to save on time. As in all ports, Doha Port's financial sustainability revolves around the number of vessels docking and the amount of cargo being processed. In essence, the greater the downtime of operations, the greater the loss of profits. Project managers must carefully weigh the scenario of completing the project faster, with additional resources, versus completing the project slower, with lesser resources. The former allows for a quicker turnaround time of port operations, thereby leading to improved business and profits, however with the cost of extra equipment. On the other hand, the latter allows for savings in construction costs, but with a slower turnaround time, penalty fees associated with delayed ship docking and loss in business and profits.

## 5   CONCLUSION

This paper developed a simulation model to mimic the berth maintenance process at Doha Port, Kuwait. The model can be used to compute minimum maintenance project duration by considering limited resources and uncertainties (e.g., equipment breakdowns and bad weather) associated with the maintenance process. The simulated production rate has been shown to be similar to the actual construction performance in Kuwait, which demonstrates the capability of the model to capture main features of berth maintenance. Uncertainty in the model tends to create project delays, thereby negatively affecting production rate and total production time. As such, adjusting resources can mitigate against possessing an unfavorable project time frame. The results of the simulation model showed that by delimiting resources and utilizing them effectively, improved construction performance can be accomplished by a contractor. Furthermore, by using the simulation model, a contractor or an owner, or a project manager can control and monitor the process of the berth maintenance project to execute the work in a minimum time as this project is a governmental project with a time limit.

In this presented case, to improve the accuracy of the model and make it more reliable, laborers/crews' working shifts timing can be included to determine the status of each worker whether they are On/Off duty. In addition, resource costs can be incorporated into the model to investigate the total cost of adopting extra equipment. This can then be compared to the costs associated with delaying ships out at sea for determining the strategy's financial feasibility. Last but not least, the tradeoff between the costs associated with adding additional resources, the saved operation downtimes and possible improvement in business and profits can be further investigated by professionals or scholars who are interested in the logistics associated with berth maintenance.

## AUTHOR BIOGRAPHIES

**RUQAYAH ALSAYED EBRAHIM** is a Master's student in the department of Civil, Environmental and Infrastructure Engineering, George Mason University. Ruqayah's current concentration is construction engineering and management. Her e-mail address is rebrahim@gmu.edu.

**SHIVANAN SINGH** is a Master's student in the department of Civil, Environmental and Infrastructure Engineering, George Mason University. Shivanan's current concentration is construction engineering and management. His e-mail address is ssingh83@gmu.edu.

**YITONG LI** is a Ph.D. candidate in the Department of Civil, Environmental & Infrastructure Engineering, George Mason University. Yitong's current research area fosuces on dynamic modeling of infrastructure restoration progress and construction simulation input modeling. Her e-mail address is yli63@gmu.edu.

**WENYING JI** is an assistant professor in the Department of Civil, Environmental & Infrastructure Engineering, George Mason University. Dr. Ji received his PhD in Construction Engineering and Management from the University of Alberta. Dr. Ji is an interdisciplinary scholar focused on the integration of advanced data analytics and complex system modeling to enhance the overall performance of infrastructure systems. His e-mail address is wji2@gmu.edu.